# Kronecker product linear exponent AR(1) correlation structures and separability tests for multivariate repeated measures

Sean L. Simpson[a,*], Lloyd J. Edwards[b], Martin A. Styner[c], and Keith E. Muller[d]

[a]*Department of Biostatistical Sciences, Wake Forest University School of Medicine, Winston-Salem, NC 27157-1063*; [b]*Department of Biostatistics, University of North Carolina at Chapel Hill, Chapel Hill, North Carolina 27599-7420*; [c]*Departments of Psychiatry and Computer Science, University of North Carolina at Chapel Hill, Chapel Hill, North Carolina 27599-7160;* [d]*Department of Health Outcomes and Policy, University of Florida, Gainesville, FL 32610-0177*

[*]Correspondence to*:* Sean L. Simpson, Department of Biostatistical Sciences, Wake Forest School of Medicine, Winston-Salem, NC 27157-1063

E-mail*:* slsimpso@wakehealth.edu

Longitudinal imaging studies have moved to the forefront of medical research due to their ability to characterize spatio-temporal features of biological structures across the lifespan. Credible models of the correlations in longitudinal imaging require two or more pattern components. Valid inference requires enough flexibility of the correlation model to allow reasonable fidelity to the true pattern. On the other hand, the existence of computable estimates demands a parsimonious parameterization of the correlation structure. For many one-dimensional spatial or temporal arrays, the *linear exponent autoregressive* (LEAR) correlation structure meets these two opposing goals in one model. The LEAR structure is a flexible two-parameter correlation model that applies in situations in which the within-subject correlation decreases exponentially in time or space. It allows for an attenuation or acceleration of the exponential decay rate imposed by the commonly used continuous-time AR(1) structure. Here we propose the Kronecker product LEAR correlation structure for multivariate repeated measures data in which the correlation between measurements for a given subject is induced by two factors. We also provide a scientifically informed approach to assessing the adequacy of a Kronecker product LEAR model and a general unstructured Kronecker product model. The approach provides useful guidance for high dimension, low sample size data that preclude using standard likelihood based tests. Longitudinal medical imaging data of caudate morphology in schizophrenia illustrates the appeal of the Kronecker product LEAR correlation structure.

## 1. Introduction

Multivariate repeated measures studies are characterized by data that have more than one set of correlated outcomes or repeated factors. Spatio-temporal data fall into this more general category since the outcome variables repeat in both space and time. Valid analysis requires accurately modeling the correlation pattern. Muller et al. (2007) and Gurka et al. (2011) showed that under-specifying the correlation structure can severely inflate test size in inference about fixed effects in the general linear mixed model. With multivariate repeated measures data, modeling the correlation pattern separately for each repeated factor has substantial advantages. Most importantly, the approach allows choosing and tuning each model separately which improves accuracy and makes model fitting easier. Furthermore, the approach inherently allows using fewer parameters than does an unstructured model. Use of the Kronecker product provides an appealing way to combine these factor-specific correlation structures into an overall correlation model. No additional parameters are needed to combine any mathematically valid correlation patterns into a valid overall pattern.

Deciding to fit a Kronecker product structure requires choosing models for each of the factors. In medical imaging, repeated measures dimensions typically have within-subject correlation decreasing exponentially in time or space. The continuous-time first-order autoregressive correlation structure, denoted AR(1), sees the most use in longitudinal settings. This model was briefly examined by Louis (1988) and is a special case of the model described by Diggle (1988). Despite its wide use, the AR(1) structure often poorly gauges within-subject correlations that decay at a slower or faster rate than required by the AR(1) model. The *linear exponent autoregressive* (LEAR) correlation model overcomes this limitation by allowing an attenuation or acceleration of the exponential decay rate imposed by the AR(1) structure (Simpson et al., 2010). Table 1 in Simpson et al (2010) contains mathematical descriptions of the LEAR and AR(1) models, as well as other stationary correlation structures that are continuous functions of distance. The focus on stationary models reflects the desire to maintain parsimony across a variety of data types. Moreover, the greater complexity of non-stationary models does not seem necessary for the limited applications of interest. It is important to note that the exponential model defined in Table 1 of Simpson et al (2010), discussed almost exclusively in the spatial statistics literature, is in fact equivalent to the continuous-time AR(1) model with $\phi = -(1/\ln \rho)$. As proposed in Simpson et al. (2010), we believe that the AR(1) and damped exponential (DE) models serve as the most relevant competitors to the LEAR structure. Special cases of both the LEAR and DE families include the AR(1), compound symmetry, and first-order moving average (MA(1)) correlation structures.

The advantages of employing a LEAR model for each component led us to consider a Kronecker product LEAR correlation structure for multivariate repeated measures data in which the correlation between measurements for a given subject is induced by two factors. We allow for an imbalance in both dimensions across subjects, i.e., an unequal numbers of observations across subjects. The LEAR model also accommodates any arbitrary spacing within a dimension. We use maximum likelihood estimation of the general linear model with Gaussian errors to illustrate the benefits of the structure. All other common estimation methods for linear and nonlinear models could also be used with a Kronecker product LEAR structure. We also provide a scientifically informed approach to testing for general and structured separability with high-dimensional data, low sample size (HDLSS) data.

For the analysis of the Kronecker product LEAR structure and tests of separability, we provide a review of separable correlation models in section 2. We then describe a motivating longitudinal imaging data example concerning schizophrenia and caudate morphology in section 3. A formal definition of the Kronecker product LEAR correlation structure is provided in section 4 along with a discussion of model estimation. In section 5 we illustrate the appeal of a Kronecker product LEAR model with the help of the caudate morphology data. We present our approach to testing for separability in HDLSS data in section 6 and illustrate its use with the caudate morphology data. Simulation studies help evaluate the approach. We conclude with a summary discussion including planned future research in section 7.

## 2. Review of separable correlation models

Galecki (1994) gave a detailed treatment of Kronecker product covariance structures, also known as separable covariance models. A covariance matrix is *separable* if and only if it can be written as $\mathbf{\Sigma} = \mathbf{\Gamma} \otimes \mathbf{\Omega}$, where $\mathbf{\Gamma}$ and $\mathbf{\Omega}$ are factor specific covariance matrices (e.g. the covariance matrices for the temporal and spatial dimensions of spatio-temporal data respectively). A key advantage of the model lies in the ease of interpretation in terms of the independent contribution of every repeated factor to the overall within-subject error covariance matrix. The model also accommodates correlation matrices with nested parameter spaces and factor specific within-subject variance heterogeneity. Galecki (1994), Naik and Rao (2001), and Mitchell et al. (2006) detailed the computational advantages of the Kronecker product covariance structure. The partial derivatives, inverse, and Cholesky decomposition of the overall covariance matrix can be performed more easily on the smaller dimensional factor specific models.

While separable covariance models are commonly used in the spatial statistics literature (Genton, 2007), they have been rarely used in multivariate longitudinal (and more generally, multivariate repeated measures) data analysis. In fact, to our knowledge, no commonly used statistical packages provide a flexible framework for implementing the structures, relegating their use to those with the appropriate programming skills. For example, SAS version 9.3 (SAS Institute, 2002) has only three Kronecker product covariance structures (unstructured matrix paired with either an unstructured, compound symmetric, or discrete-time AR(1) matrix). Given the advantages of separable models, extending software to allow their general implementation is important for researchers in a variety of areas. For example, longitudinal group-randomized controlled trials often have within-group correlations (e.g., by school for youth based studies) and within-subject longitudinally induced correlations (Komro et al., 2007). Such data can be well modeled by taking the Kronecker product of a compound symmetric and LEAR correlation structure.

Limitations of separable models have been noted by various authors. Most importantly, as mentioned by Cressie and Huang (1999), patterns of interaction among the various factors cannot be modeled when utilizing a Kronecker product structure. Galecki (1994), Huizenga et al. (2002), and Mitchell et al. (2006) all noted that a lack of identifiability can result with such a model. The indeterminacy stems from the fact that if $\mathbf{\Sigma} = \mathbf{\Gamma} \otimes \mathbf{\Omega}$ is the overall within-subject error covariance matrix, $\mathbf{\Gamma}$ and $\mathbf{\Omega}$ are not unique since for $a \neq 0$, $a\mathbf{\Gamma} \otimes (1/a)\mathbf{\Omega} = \mathbf{\Gamma} \otimes \mathbf{\Omega}$. However, this nonidentifiability can be fixed by rescaling one of the factor specific covariance matrices so that one of its diagonal nonzero elements is equal to 1. With homogeneous variances, the rescaled matrix is a correlation matrix. It is also important to note that within a given subject all factors must have *consistently-spaced* measurements. In the context of spatio-temporal data this means that at each time point a given subject must have the same number of measurements taken at the same spatial locations.

Several tests have been developed to determine the validity of assuming a separable covariance model. General (pure) tests use unstructured null and alternative hypothesis matrices. Shitan and Brockwell (1995) constructed an asymptotic chi-square test for general separability. Likelihood ratio tests for general separability were derived by Lu and Zimmerman (2005), Mitchell et al. (2006), and Roy and Khattree (2003). Fuentes (2006) developed a general test for separability of a spatio-temporal process utilizing spectral methods.

Structure-specific tests of separability have particular structure assumed for the null hypothesis but generally not for the alternative hypothesis. Structured tests of separability have been proposed by Roy and Khattree (2005a, 2005b) and Roy and Leiva (2008). Roy and Khattree (2005a) derived a test for the case with one factor matrix being compound symmetric and the other unstructured. Roy and Khattree (2005b) developed a test for when one factor specific matrix has the discrete-time AR(1) structure and the other is unstructured. The test of Roy and Leiva (2008) requires either a compound symmetric or discrete-time AR(1) structure for the factor specific matrices. Simpson (2010) developed an adjusted likelihood ratio test of two-factor separability for unbalanced multivariate repeated measures data. The approach can be generalized to factor specific matrices of any structure.

All of the authors just mentioned noted that none of the separability tests developed thus far can handle HDLSS data due to the nonexistence of an estimate for an unstructured covariance fit (the alternative hypothesis). We provide a scientifically informed approach to conducting useful tests of separability in the presence of HDLSS, a common problem in medical imaging and various kinds of "-omics" data.

## 3. Example: schizophrenia and caudate morphology

Disabling impairments in the perception or expression of reality characterize schizophrenia. Pathological changes in brain morphology in schizophrenics may be progressive and associated with clinical outcome. Much recent work has focused on the effect of antipsychotic drugs on brain morphology. The caudate, an important part of the brain's learning and memory system, has been one target of the drugs.

Our data come from longitudinal MRI scans of the left caudate for 240 schizophrenia patients and 56 controls. The surface of each object was parameterized via the m-rep method as described in Styner and Gerig (2001). The caudate shape was determined as a 3 x 7 grid of mesh points (see Figure 1). Data were reduced to one outcome measure: *radius* in cm as a measure of local object width (21 locations per caudate). The distance between two radii for a given subject was calculated as the mean Euclidian distance over all images. Scans were taken up to 47 months post-baseline with the median and maximum number of scans per subject being 3 and 7 respectively. The schizophrenia patients were randomized to either haloperidol (a conventional antipsychotic) or olanzapine (an atypical antipsychotic). The two groups were combined into one treatment group in our analysis in order to avoid undermining ongoing research. The other covariates of interest were age, gender, and race. Preliminary analyses showed that the shape of the caudate, and thus the radii, differs significantly at baseline between schizophrenics and controls. The study hypothesized that the neuroprotective effect of the drugs would lead to no overall differences in shape between the patients and controls.

## 4. Kronecker product lear correlation structure

**4.1 *Definition***

With the assumptions of covariance model separability and homoscedasticity, an equal variance Kronecker product structure has great appeal. The overall within-subject error covariance matrix is defined as $\mathbf{\Sigma}_i = \sigma^2 \mathbf{\Gamma}_i \otimes \mathbf{\Omega}_i$ for the $i^{th}$ subject or independent sampling unit. The formulation has several advantages. The reduction in the number of parameters leads to computational benefits. The model is also identifiable since $\mathbf{\Gamma}_i$ and $\mathbf{\Omega}_i$ will necessarily be correlation matrices. When heteroscedasticity is present, $\sigma^2$ can be thought of as an aggregate variance parameter for the two factors.

Suppose $\boldsymbol{y}_i$ is a $t_i s_i \times 1$ vector of $t_i s_i$ observations (e.g., $t_i$ temporal measurements and $s_i$ spatial measurements) on the $i^{th}$ subject $i \in \{1, \ldots, N\}$. Here $\mathcal{C}(y_{ijl}, y_{ikl}) = \rho_{i\gamma;jk}$ and $\mathcal{C}(y_{ijl}, y_{ijm}) = \rho_{i\omega;lm}$ represent the temporal (or factor 1) and spatial (or factor 2) correlations respectively, for $\mathcal{C}(\,\cdot\,)$ the correlation operator. Then for $\mathbf{\Gamma}_i = \{\rho_{i\gamma;jk}\}$ (the temporal/factor 1 correlation matrix) and $\mathbf{\Omega}_i = \{\rho_{i\omega;lm}\}$ (the spatial/factor 2 correlation matrix), the factor specific *linear exponent autoregressive* (LEAR) correlation structures are

$$\rho_{i\gamma;jk} = \mathcal{C}(y_{ijl}, y_{ikl}) = \begin{cases} \rho_\gamma^{d_{t;\min} + \delta_\gamma[(d(t_{ijl}, t_{ikl}) - d_{t;\min})/(d_{t;\max} - d_{t;\min})]} & j \neq k \\ 1 & j = k \end{cases}, \tag{1}$$

$$\rho_{i\omega;lm} = \mathcal{C}(y_{ijl}, y_{ijm}) = \begin{cases} \rho_\omega^{d_{s;\min} + \delta_\omega[(d(s_{ijl}, s_{ijm}) - d_{s;\min})/(d_{s;\max} - d_{s;\min})]} & l \neq m \\ 1 & l = m \end{cases}. \tag{2}$$

The Kronecker product LEAR correlation structure is

$$\mathbf{\Sigma}_i = \mathbf{\Gamma}_i \otimes \mathbf{\Omega}_i, \tag{3}$$

where $d(t_{ijl}, t_{ikl})$ and $d(s_{ijl}, s_{ikl})$ are the distances between measurement times and locations respectively. In turn $(d_{t;\min}, d_{s;\min})$ and $(d_{t;\max}, d_{s;\max})$ are computational *constants* equal to the minimum and maximum number of temporal and spatial distance units across all subjects. Parameters $\rho_\gamma$ and $\rho_\omega$ are the correlations between observations separated by one unit of time and distance respectively, and $\delta_\gamma$ and $\delta_\omega$ are the decay speeds. We assume $0 \leq \rho_\gamma, \rho_\omega < 1$ and $0 \leq \delta_\gamma, \delta_\omega$. The $(d_{t;\min}, d_{s;\min})$ and $(d_{t;\max}, d_{s;\max})$ constants allow the model to adapt to the data and scale distance such that the multiplier of the decay speeds $\delta_\gamma$ and $\delta_\omega$, $(d(t_{ijl}, t_{ikl}) - d_{t;\min})/(d_{t;\max} - d_{t;\min})$ and $(d(s_{ijl}, s_{ijm}) - d_{s;\min})/(d_{s;\max} - d_{s;\min})$, is between 0 and 1 for computational purposes. One could also consider tuning the constants if necessary to address, for example, convergence issues. Simpson et al. (2010) gave details on setting the distance constants. Ensuring that the factor specific matrices $\mathbf{\Gamma}_i$ and $\mathbf{\Omega}_i$ are positive definite (as discussed in Simpson et al. (2010)) is sufficient for ensuring the positive definiteness of $\mathbf{\Sigma}_i$ (Theorem 7.10, Schott, 1997).

Graphical depictions of the Kronecker product LEAR structure help to provide insight into the types of correlation patterns that can be modeled. A correlation pattern in which both of the factor specific matrices (e.g. spatial and temporal matrices) have decay rates slower than that of the AR(1) model is illustrated in Figure 2A. Figures 2B and 2C exhibit patterns with dual AR(1) and faster than AR(1) decay rates respectively.

Given the advantages of the Kronecker product covariance model, we believe that it has been underutilized in practice. As alluded to in section 2, the model has great computational properties and simplifies interpretation. It also reduces the dimension of the calculations, sometimes drastically (e.g., having two $30 \times 30$ matrices vs. a $900 \times 900$ matrix), while allowing complex factor-specific correlation structures. These inherent qualities make the Kronecker product covariance model an appealing solution to the High Dimension, Low Sample Size problem so common in medical imaging and various kinds of "-omics" data. Modeling the factor specific matrices with the LEAR structure is especially attractive due to the increased flexibility, parsimony, and numerical stability resulting from this combination.

Here we adopt the technique of modeling the correlation and variance structures separately as done in the approaches of Fan et al. (2007) and others. We focus on modeling the correlation and henceforth assume an equal variance structure for the application of interest. Thus,

$$\mathbf{\Sigma}_i = \sigma^2(\mathbf{\Gamma}_i \otimes \mathbf{\Omega}_i). \tag{4}$$

## 4.2 *Estimation*

The Kronecker product LEAR structure can be imbedded within various modeling and estimation methods. The best approach may vary by context. With linear structured factor specific matrices, the noniterative approach of Werner et al. (2008) has appeal. However, the approach is not appropriate for the LEAR structure given its nonlinear nature. Naik and Rao (2001), Huizenga et al. (2002), Lu and Zimmerman (2005), Mitchell et al. (2006), Roy and Khattree (2003, 2005a, 2005b), and Roy and Leiva (2008) all use maximum likelihood methods for parameter estimation in a Kronecker product model. However, none of their approaches allow for data that are unbalanced in both dimensions. As noted by Edwards et al. (2008) and others, the Kenward-Roger approach with REML estimation is preferable for small sample estimation and inference. Non- and semiparametric approaches, like those of Wang (2003) and Fan et al. (2007), may prove beneficial for non-Gaussian data. The Kronecker product LEAR model may also serve as a plausible working correlation structure in a generalized estimating equation (GEE) framework. We focus on Gaussian data with moderately large sample sizes, and leave the examination of the Kronecker product model in other contexts to future work.

Here we consider maximum likelihood (ML) estimation of the general linear model where the Gaussian errors have a Kronecker product LEAR correlation structure. We allow for an imbalance in both dimensions of the data. ML estimation is used to allow the model to be amenable to likelihood ratio tests of separability like those noted in section 2.

Consider the following general linear model for multivariate repeated measures data with the Kronecker product LEAR correlation structure:

$$\boldsymbol{y}_i = \boldsymbol{X}_i \boldsymbol{\beta} + \boldsymbol{e}_i \tag{5}$$

where again $\boldsymbol{y}_i$ is a $t_i s_i \times 1$ vector of $t_i s_i$ observations (e.g., $t_i$ temporal measurements and $s_i$ spatial measurements) on the $i^{th}$ subject $i \in \{1, \ldots, N\}$, $\boldsymbol{\beta}$ is a $q \times 1$ vector of fixed and unknown population parameters, $\boldsymbol{X}_i$ is a $t_i s_i \times q$ fixed and known design matrix corresponding to the fixed effects, and $\boldsymbol{e}_i$ is a $t_i s_i \times 1$ vector of random error terms. We assume $\boldsymbol{e}_i \sim N_{t_i s_i}(\mathbf{0}, \boldsymbol{\Sigma}_i = \sigma^2[\boldsymbol{\Gamma}_i \otimes \boldsymbol{\Omega}_i])$ is independent of $\boldsymbol{e}_{i'}$ for $i \neq i'$. It follows that $\boldsymbol{y}_i \sim N_{t_i s_i}(\boldsymbol{X}_i\boldsymbol{\beta}, \sigma^2[\boldsymbol{\Gamma}_i \otimes \boldsymbol{\Omega}_i])$ is independent of $\boldsymbol{y}_{i'}$ for $i \neq i'$, where $\boldsymbol{\Gamma}_i$ and $\boldsymbol{\Omega}_i$ are defined in equations 1 and 2.

Setting $\boldsymbol{\Sigma}_i = \sigma^2[\boldsymbol{\Gamma}_i(\boldsymbol{\tau}_\gamma) \otimes \boldsymbol{\Omega}_i(\boldsymbol{\tau}_\omega)]$, where $\boldsymbol{\tau} = \{\boldsymbol{\tau}_\gamma; \boldsymbol{\tau}_\omega\} = \{\delta_\gamma, \rho_\gamma; \delta_\omega, \rho_\omega\}$, the log-likelihood function of the parameters given the data under the model is:

$$
\begin{aligned}
&l(\boldsymbol{y}; \boldsymbol{\beta}, \sigma^2, \boldsymbol{\tau}) \qquad (6)\\
&= -\frac{n}{2}\ln(2\pi) - \frac{1}{2}\sum_{i=1}^{N}\ln|\sigma^2\boldsymbol{\Gamma}_i \otimes \boldsymbol{\Omega}_i| - \frac{1}{2\sigma^2}\sum_{i=1}^{N}\boldsymbol{r}_i(\boldsymbol{\beta})'(\boldsymbol{\Gamma}_i \otimes \boldsymbol{\Omega}_i)^{-1}\boldsymbol{r}_i(\boldsymbol{\beta})\\
&= -\frac{n}{2}\ln(2\pi) - \frac{1}{2}\sum_{i=1}^{N}\big(t_i s_i \ln(\sigma^2) + \ln|\boldsymbol{\Gamma}_i \otimes \boldsymbol{\Omega}_i|\big) - \frac{1}{2\sigma^2}\sum_{i=1}^{N}\boldsymbol{r}_i(\boldsymbol{\beta})'(\boldsymbol{\Gamma}_i \otimes \boldsymbol{\Omega}_i)^{-1}\boldsymbol{r}_i(\boldsymbol{\beta}),
\end{aligned}
$$

where $n = \sum_{i=1}^{N} t_i s_i$ and $\boldsymbol{r}_i(\boldsymbol{\beta}) = \boldsymbol{y}_i - \boldsymbol{X}_i\boldsymbol{\beta}$. The ML estimates are derived following the approach employed in Simpson et al. (2010). After profiling $\sigma^2$ out of the likelihood, the profile log-likelihood is given by

$$
\begin{aligned}
l_p(\boldsymbol{y}; \boldsymbol{\beta}, \boldsymbol{\tau}) = -\frac{n}{2}\ln(2\pi) - \qquad (7)\\
\frac{1}{2}\sum_{i=1}^{N}\ln|\boldsymbol{\Gamma}_i \otimes \boldsymbol{\Omega}_i| - \frac{1}{2}n\ln\left[\sum_{i=1}^{N}\boldsymbol{r}_i(\boldsymbol{\beta})'(\boldsymbol{\Gamma}_i^{-1} \otimes \boldsymbol{\Omega}_i^{-1})\boldsymbol{r}_i(\boldsymbol{\beta})\right] + \frac{1}{2}n\ln n - \frac{n}{2}
\end{aligned}
$$

To avoid computational issues it is best to use the equality

$$\ln|\boldsymbol{\Gamma}_i \otimes \boldsymbol{\Omega}_i| = s_i \ln|\boldsymbol{\Gamma}_i| + t_i \ln|\boldsymbol{\Omega}_i|$$

in case $|\boldsymbol{\Gamma}_i \otimes \boldsymbol{\Omega}_i|$ is close to zero.

The ML estimates of the model parameters may be computed with the Newton-Raphson algorithm which requires the first and second partial derivatives of the profile log-likelihood. The derivations of the first partial derivatives are available from the authors. The second partial derivatives of the parameters, which are employed to determine the asymptotic variance-correlation matrix of the estimators, may be approximated by finite difference formulas. The derivative approximations are detailed in Abramowitz and Stegun (1972) and Dennis and Schnabel (1983). The 15 analytic second derivatives can be derived explicitly as in Simpson et al. (2010). However, the approximations have proven very accurate.

After getting the estimates of $\boldsymbol{\beta}$ and $\boldsymbol{\tau}$ utilizing the Newton-Raphson algorithm, an estimate of $\sigma^2$ is calculated by substituting the estimates into $\widehat{\sigma}^2{}_{ML}(\boldsymbol{\beta}, \boldsymbol{\tau}) = n^{-1}\sum_{i=1}^{N}\boldsymbol{r}_i(\boldsymbol{\beta})'(\boldsymbol{\Gamma}_i^{-1} \otimes \boldsymbol{\Omega}_i^{-1})\boldsymbol{r}_i(\boldsymbol{\beta})$, which is the expression resulting from the initial profiling of $\sigma^2$ out of the likelihood. An estimator of the variance for $\widehat{\sigma}^2{}_{ML}(\boldsymbol{\beta}, \boldsymbol{\tau})$, assuming that $\boldsymbol{\beta}$ and $\boldsymbol{\tau}$ are known, is then

$$\widehat{\mathcal{V}}\left[\widehat{\sigma}^2{}_{ML}(\boldsymbol{\beta}, \boldsymbol{\tau})\right] = 2\widehat{\sigma}^4/n. \tag{8}$$

The derivation of this estimator is available from the authors. A SAS IML (SAS Institute, 2002) program implementing this estimation procedure for the general linear model with a Kronecker product LEAR correlation structure is also available upon request.

A complication that may arise when implementing the Kronecker product LEAR correlation model is that the proposed estimation method can produce negative variance estimates for the correlation parameters. This may occur for the parameters of either one or both of the factor specific matrices when there is a faster decay rate than that imposed by the AR(1) model coupled with a "small" $\rho_\gamma$ and/or $\rho_\omega$. The instability of the second order derivatives of the objective function resulting from the small, quickly decaying correlation(s) leads to this problem.

An alternate approach would be to implement an estimation method which uses only first order derivatives such as a quasi-Newton procedure. An efficient modification of Powell's (1978a, 1978b, 1982a, 1982b) Variable Metric Constrained WatchDog (VMCWD) algorithm is often used. A quadratic programming subroutine updates and downdates the Cholesky factor as detailed by Gill et al. (1984). However, quasi-Newton approaches generally have worse stability and convergence properties than the Newton-Raphson method. Another approach is to recognize this complication as a diagnostic tool. Since a correlation matrix of this nature is approximately equal to the identity matrix, an independence model may be the best fit for the factor specific structure in this situation.

**5. Data applications and results**

We model the caudate data discussed in section 3 with the general linear model for multivariate repeated measures data defined in section 4.2 (all modeling assumptions were assessed and met as discussed in Muller and Stewart, 2006). The initial full mean model is as follows:

$$\boldsymbol{y}_i = \beta_0 + \beta_1 \boldsymbol{X}_{i,\text{trt}} + \beta_2 \boldsymbol{X}_{i,\text{age}} + \beta_3 \boldsymbol{X}_{i,\text{gen}} + \beta_4 \boldsymbol{X}_{i,\text{af_race}} + \beta_5 \boldsymbol{X}_{i,\text{oth_race}} + \boldsymbol{e}_i. \tag{9}$$

The $\log_2$(radius) values for each of the $s_i = s = 21$ locations (spatial factor) and $t_i$ images (temporal factor) for each subject are contained in $\boldsymbol{y}_i$ $(t_i{\cdot}21 \times 1)$. The vectors $\boldsymbol{X}_{i,\text{trt}}$, $\boldsymbol{X}_{i,\text{gen}}$, $\boldsymbol{X}_{i,\text{af_race}}$ and $\boldsymbol{X}_{i,\text{oth_race}}$ indicate the treatment group (patients and controls), gender, and race (African-American, Other, and White--reference group) of the $i^{\text{th}}$ subject respectively. The ages at baseline are contained in $\boldsymbol{X}_{i,\text{age}}$.

We first assume a separable covariance and model the temporal and spatial factor specific correlations of the model errors with continuous-time AR(1), DE, and LEAR structures in order to assess the best model via the AIC. A test of separability for the data example is provided in Section 6. Table 1 contains the AIC values for all nine possible correlation model fits with the initial full mean model. Modeling both the temporal and spatial correlations with the LEAR structure provides the best model fit of the nine combinations. The BIC corroborates these differences in fits. The resulting parameter estimates and p-values (based on the residual approximation of the $F$-test for a Wald statistic) associated with each of the covariates are presented in Table 2 for three of the correlation model fits: LEAR $\otimes$ LEAR, AR(1) $\otimes$ AR(1), and DE $\otimes$ DE. Though, for our particular example, there is no difference in fixed effect (mean model) inference among the models, the covariate p-values for the better fitting LEAR $\otimes$ LEAR model are uniformly larger for $\beta_2$ (age), $\beta_3$ (gender), and $\beta_4$ and $\beta_5$ (race). Thus, these results illuminate the difference in fixed effect inference that could occur in a more marginal context. The better fit of the Kronecker product LEAR structure also gives more confidence in the results of the analysis.

We continue the analysis employing the Kronecker product LEAR correlation model. In order to obtain a parsimonious model, the full model defined in equation 9 is reduced via backward selection with $\alpha = 0.20$. At each reduction step the covariance parameters are re-estimated and the fixed effect covariate with the largest p-value is removed if it is non-significant at the $\alpha = 0.20$ level based on the residual approximation of the $F$-test for a Wald statistic. The final model after reduction is

$$\boldsymbol{y}_i = \beta_0 + \boldsymbol{e}_i. \tag{10}$$

Thus, as expected, there is no evidence of a difference in caudate shape between the treated schizophrenics and the controls when taking into account all images taken over time. To ensure that the Kronecker product LEAR correlation model still provides the better fit for the final model, we again model the temporal and spatial factor specific correlations of the model errors with the continuous-time AR(1), DE, and LEAR structures. Table 1 contains the AIC values for the final model fits. We see that the Kronecker product LEAR correlation model remains the better correlation structure for the final data model. The BIC corroborates the differences in fits.

The residual variance estimate and correlation parameter estimates of the Kronecker product LEAR structure (defined in equation 4) for the final data model are given in Table 3. Graphical depictions of these estimates are exhibited in Figure 3, which show the observed vs. predicted correlation patterns as a function of the months between images and millimeters between radii respectively, starting with the minimum temporal and spatial distances for the data. As evidenced by Figure 3A, the temporal factor specific LEAR correlation structure is able to model a correlation function in which the correlation remains high regardless of how far apart in time the images are taken. The fact that the correlation estimates in the time dimension are close to unity might be considered a problem if the presence of a unit root is expected. This would also present a problem for the competing DE and AR(1) factor specific models. While much work has been done on the development of unit root tests for time series data (Im et al., 2003; Baltagi et al., 2007; Moon and Perron 2012; Westerlund and Larsson, 2012; Lin, 2013), to our knowledge, no tests have been developed for our context. That is, none are applicable to unbalanced, inconsistently-spaced multivariate repeated measures data modeled with a Kronecker product covariance structure. A test for a unit root might be useful, but developing one is beyond the scope of the present work. The spatial correlations, shown in Figure 3B, are modest for radii that are close, and then decay slowly toward zero as they become farther apart. The predicted correlation curve appears to slightly overestimate the spatial correlations for small distances. This may be due to the restriction $0 \leq \rho_\omega < 1$ since the model cannot accurately incorporate the negative correlations. One solution may be to add an offset parameter to the model in order to account for negative correlations, i.e.,

$$\rho_{0_\omega} + \rho_\omega^{d_{s;\min} + \delta_\omega[(d(s_{ijl}, s_{ijm}) - d_{s;\min})/(d_{s;\max} - d_{s;\min})]},$$

where

$$-1 - \rho_\omega^{d_{s;\min} + \delta_\omega[(d(s_{ijl}, s_{ijm}) - d_{s;\min})/(d_{s;\max} - d_{s;\min})]} < \rho_{0_\omega} < 1 - \rho_\omega^{d_{s;\min} + \delta_\omega[(d(s_{ijl}, s_{ijm}) - d_{s;\min})/(d_{s;\max} - d_{s;\min})]}.$$

An examination of this approach, and others, will be left for future research.

The predicted correlation as a function of both factors is exhibited in Figure 4. This illustration of the predicted overall within-subject correlation function again displays the slow spatial decay pattern and the near constant temporal pattern. The utility of the Kronecker product correlation model lies in both the flexibility of the factor specific models as well as the interpretability stemming from the Kronecker product structure.

## 6. Likelihood ratio tests of separability

### 6.1 *Definition*

We examine both a structured and general test of separability to illustrate our approach. We consider the following structured likelihood ratio test of separability for the KP LEAR model detailed in section 4:

$$H_0\colon \boldsymbol{\Sigma}_i = \sigma^2 \boldsymbol{\Gamma}_i \otimes \boldsymbol{\Omega}_i;\ \boldsymbol{\Gamma}_i, \boldsymbol{\Omega}_i \text{ LEAR vs. } H_1\colon \boldsymbol{\Sigma}_i \text{ unstructured, positive definite (PD).} \quad (11)$$

Following the notation from section 4, the standard likelihood ratio is given by

$$\Lambda = \frac{\max_{H_0} L}{\max_{H_1} L} = \frac{\exp\left\{ -\frac{1}{2\widehat{\sigma}^2} \sum_{i=1}^{N} \boldsymbol{r}_i(\widehat{\boldsymbol{\beta}}_0)'(\widehat{\boldsymbol{\Gamma}}_i^{-1} \otimes \widehat{\boldsymbol{\Omega}}_i^{-1}) \boldsymbol{r}_i(\widehat{\boldsymbol{\beta}}_0) \right\} \prod_{i=1}^{N} \widehat{\sigma}^{-t_i s_i} |\widehat{\boldsymbol{\Gamma}}_i|^{-s_i/2} |\widehat{\boldsymbol{\Omega}}_i|^{-t_i/2}}{\exp\left\{ -\frac{1}{2} \sum_{i=1}^{N} \boldsymbol{r}_i(\widehat{\boldsymbol{\beta}}_1)' \widehat{\boldsymbol{\Sigma}}_i^{-1} \boldsymbol{r}_i(\widehat{\boldsymbol{\beta}}_1) \right\} \prod_{i=1}^{N} (2\pi)^{-t_i s_i/2} |\widehat{\boldsymbol{\Sigma}}_i|^{-1/2}}. \quad (12)$$

Under regularity conditions, $-2\ln\Lambda$ is asymptotically distributed as a $\chi^2_\nu$ random variable. The associated degrees of freedom parameter $\nu$ is given by

$$\nu = \max_i \left( \frac{t_i s_i (t_i s_i + 1)}{2} \right) - 5. \quad (13)$$

Here we use the adjusted LRT (aLRT) of Simpson (2010), namely $-2\ln\Lambda \approx k\chi^2_\nu$, where

$$k = N \Big/ \left( N - \max_i (t_i s_i) \right) \quad (14)$$

and conduct several tests of "marginal KP LEARness" since the number of observations per subject precludes conducting an overall aLRT using all of the data.

We also consider the following general likelihood ratio test of separability:

$$H_0\colon \boldsymbol{\Sigma}_i = \boldsymbol{\Gamma}_i \otimes \boldsymbol{\Omega}_i;\ \boldsymbol{\Gamma}_i, \boldsymbol{\Omega}_i \text{ unstructured, PD vs. } H_1\colon \boldsymbol{\Sigma}_i \text{ unstructured, PD.} \quad (15)$$

We again employ the aLRT as defined in equations 12 and 14, with the associated degrees of freedom $\nu$ now given by

$$\nu = \max_i \left( \frac{t_i s_i (t_i s_i + 1)}{2} \right) - \max_i \left( \frac{t_i (t_i + 1)}{2} + \frac{s_i (s_i + 1)}{2} - 1 \right). \quad (16)$$

As stated previously, we would like to provide a scientifically informed approach to assessing the appropriateness of a separable model with high-dimensional data. To do this we will conduct $c$ marginal aLRT tests using subsets of the data corresponding to diagonal blocks of the covariance matrices. The approach covers a large fraction of the covariance space and is especially useful in situations with most of the information contained along diagonal blocks (i.e., correlation dies out along off-diagonals), which is the case for the data discussed in section 3. For balanced data, each diagonal subset will contain data from all subjects. However, for unbalanced data, subsets should either be chosen so that the same number of subjects are used in each test or some weighting of the $c$ tests should be considered based on the number of subjects in each. We take the former approach in the simulation studies and data application that follow. After the $c$ aLRT tests are conducted, a false discovery rate correction is applied that controls for multiple testing given dependent tests (Benjamini and Yekateuli, 2001) and significance of the overall test is declared if any of the $c$ p-values is significant. The dependencies among covariance parameters in samples from a multivariate Gaussian population were given by Wishart (1928).

This marginal testing approach can be thought of as a generalization of the more formalized framework of Molenberghs et al. (2011). They presented a pseudo-likelihood based method to partition prohibitively large data sets into $M$ sub-samples, analyze each partition member, and combine the results across partitions for parameter estimation. More formally, the full sample is broken into $M$ sub-samples of size $n_m$, where $m \in \{1, \ldots, M\}$. The pseudo-likelihood for sample $m$ is

$$pl_m(\boldsymbol{\theta}_m) = \sum_{i=1}^{n_m} l(\boldsymbol{y}_{mi}|\boldsymbol{\theta}_m), \tag{17}$$

where $l(\,\cdot\,)$ is the likelihood that would be considered if the $m^{\text{th}}$ sub-sample were the entire data set. In our case, $l(\,\cdot\,) = l(\boldsymbol{y}_i; \boldsymbol{\beta}, \sigma^2, \boldsymbol{\tau})$ from equation 6 for the sub-sample under $H_0$ for the structured test and $l(\,\cdot\,) = l(\boldsymbol{y}_i; \boldsymbol{\beta}, \boldsymbol{\Gamma}_i, \boldsymbol{\Omega}_i)$ under $H_0$ for the general test. Under $H_1$, $l(\,\cdot\,) = l(\boldsymbol{y}_i; \boldsymbol{\beta}, \boldsymbol{\Sigma}_i)$ for both the structured and general test. Extending their approach directly into our context would involve averaging test statistics or p-values across sub-samples (as they do for estimation of $\boldsymbol{\theta}$) and drawing inference from the averaged value. Our approach is slightly different in that we test for marginal separability for each of the $M$ sub-samples and draw inference based on the number of these $M$ tests that are significant. Our method provides robustness against outlying sub-sample p-values.

## 6.2 *Simulation Studies*

### 6.2.1 *Structured Tests*

To assess the empirical performance of the structured aLRT in equation 11, we conducted two simulation studies. The objective of the first study was to assess how much information is lost in taking the diagonal subset approach. We simulated multivariate repeated measures data with a $16 \times 16$ covariance matrix under $H_0$ (two $4 \times 4$ factor specific matrices) and $H_1$and then conducted 1) an aLRT using all of the data, and 2) four tests using diagonal $2 \times 2$ subsets of the factor specific matrices following the testing procedure delineated in the previous section (subset aLRT). The data were generated under $H_0$ with $\boldsymbol{\rho} = [\,\rho_\gamma \quad \rho_\omega\,]' = [\,0.8 \quad 0.8\,]'$, $\boldsymbol{\delta} = [\,\delta_\gamma \quad \delta_\omega\,]' = [\,(d_{t;\max} - d_{t;\min})/4 \quad (d_{s;\max} - d_{s;\min})/4\,]'$, $\sigma^2 = 1$, and two-unit distance intervals for both factors (space and time). The data were generated under $H_1$ based on Theorem 10.13 of Muller and Stewart (2006). Simulated test size and power at the $\alpha = 0.05$ level was examined for tests 1) and 2) with sample sizes of $N = 40, 80, 120, 160,$ and 200. Without lose of generality, the mean model was set to $\beta = 0$ (one group with mean 0) (Simpson, 2010). Each simulation consisted of 5,000 realizations.

Table 4 shows the results of this first simulation study for the structured tests. It contains the simulated test sizes and power for the full aLRT and subset aLRT approaches. For the subset approach, the table also shows the test size and power by the number (out of $c = 4$) of significant p-values required for overall significance to be declared. There is very slight test size inflation for the subset aLRT when one significant p-value is required. Test size is well controlled when more than one significant p-value is required as it is for the full data aLRT. Additionally, there is no loss in power for the subset aLRT as compared with the full data aLRT. Although the subset method is for situations in which a full data test is infeasible, this comparison gives us confidence that little information is lost when taking our subset approach.

The objective of the second simulation study was to assess the type I error rate and power of the subset aLRT approach in a higher-dimensional setting in which current separability tests are unsuitable due to the nonexistence of an estimate for an unstructured covariance fit. Our simulations were aimed to mimic the schizophrenia and caudate morphology data discussed in section 3. To do this, we again generated data under $H_0$ with $\boldsymbol{\rho} = [\,\rho_\gamma \quad \rho_\omega\,]' = [\,0.8 \quad 0.8\,]'$, $\boldsymbol{\delta} = [\,\delta_\gamma \quad \delta_\omega\,]' = [\,(d_{t;\max} - d_{t;\min})/4 \quad (d_{s;\max} - d_{s;\min})/4\,]'$, and $\sigma^2 = 1$, and under $H_1$ based on Theorem 10.13 of Muller and Stewart (2006). To mimic the example, the data generated were unbalanced with $\max_i(s_i) = s = 21$, $\max_i(t_i) = 7$, and $\text{med}_i(t_i) = 3$. There were $(t_i \cdot s_i)\,\epsilon$ $[21, 147]$ observations per subject, each at two-unit distance intervals. Ten subset tests of $2 \times 2$ diagonal blocks of the spatial matrix were conducted using the entire $7 \times 7$ temporal matrix $\boldsymbol{\Gamma}_i \otimes \boldsymbol{\Omega}$. Ten tests occur since there are ten $2 \times 2$ diagonal blocks in the $21 \times 21$ spatial matrix with 1 diagonal element omitted. Given the imbalance in temporal measurements, the approach ensures that the same number of subjects are used in each subtest. Simulated test size and power at the $\alpha = 0.05$ level was examined for sample sizes of $N = 240, 280,$ and 320.

The results of the second simulation study for the structured tests are shown in Table 5. It contains the simulated test size and power for the subset aLRT by the number (out of $c = 10$) of significant p-values required for overall significance to be declared. There is severe test size inflation for all sample sizes when only 1 significant p-value is required. However, the test size becomes controlled for $N = 280$ and 320 when 2 or more significant values are required, and for $N = 240$ when 3 or more are required. Regardless of the number of significant subtest p-values required for overall significance, the test remains extremely powerful.

6.2.2 *General Tests*

To assess the empirical performance of the unstructured aLRT in equation 15, we conducted the same two simulation studies detailed in the previous subsection for the structured case. Here the data were generated under $H_0$ and $H_1$ based on Theorem 10.13 of Muller and Stewart (2006). Table 4 exhibits the results of the first simulation study for the general tests assessing information loss when taking the diagonal subset approach. Test size is well controlled for the subset aLRT across all conditions as it is for the full data aLRT. Moreover, there is minimal to no loss in power for all sample sizes with the subset aLRT when one significant p-value is required. However, the power loss increases as the number of significant p-values required increases, with this effect mitigated at larger sample sizes.

Table 5 displays the results of the second simulation study for the general tests assessing the type I error rate and power of the subset aLRT approach in the higher-dimensional setting. Severe test size inflation occurs for all sample sizes when only 1 significant subtest p-value is required and remains above the $\alpha = 0.05$ level until 6 or more (out of $c = 10$) significant values are required for $N = 280$ and 320 and 6 or more are required for $N = 240$. As with the structured test, the unstructured subset aLRT remains extremely powerful for all parameter combinations.

### 6.3 *Test of Separability for Data Example*

As evidenced by the analysis in section 5, the Kronecker product LEAR model provides a good fit to the spatial and temporal correlations in the data. However, the validity of the separable assumption should be assessed as there may be space $\times$ time interactions which cannot be modeled with the Kronecker structure. In order to test separability, Simpson (2010) was forced to reduce the data by picking four (out of the 21) representative spatial locations to accommodate the dimensions of the data. Here we apply our structured subset aLRT approach detailed in section 6.1 by conducting 10 subset tests utilizing the entire $7 \times 7$ temporal matrix $\mathbf{\Gamma}_i \otimes \mathbf{\Omega}$ $2 \times 2$ diagonal blocks of the spatial matrix as in the second simulation study. For a significance level of $\alpha = 0.05$ and, based on the simulations conducted in section 6.2.1, declare significance if 3 or more (out of $c = 10$) of the subtest p-values are significant. The null hypothesis is rejected since more than 3 of the tests had significant p-values, which implies that the assumption of separability appears invalid in this case. In order to gain insight into this finding we examine the following estimates (each multiplied by 100) for 4 (of the 21) spatial locations and the 3 time points for subject $i = 4$ $(t_4 = 3, s = 4)$ (also in Simpson, 2010):

$$\widehat{\sigma}^2\widehat{\mathbf{\Gamma}}_4\otimes\widehat{\mathbf{\Omega}}=\left(\begin{array}{cccc|cccc|cccc}
1.36 & 0.49 & 0.42 & 0.24 & 1.08 & 0.39 & 0.34 & 0.19 & 1.07 & 0.38 & 0.33 & 0.19\\
0.49 & 1.36 & 0.28 & 0.33 & 0.39 & 1.08 & 0.22 & 0.26 & 0.38 & 1.07 & 0.22 & 0.26\\
0.42 & 0.28 & 1.36 & 0.18 & 0.34 & 0.22 & 1.08 & 0.14 & 0.33 & 0.22 & 1.07 & 0.14\\
0.24 & 0.33 & 0.18 & 1.36 & 0.19 & 0.26 & 0.14 & 1.08 & 0.19 & 0.26 & 0.14 & 1.07\\
\hline
1.08 & 0.39 & 0.34 & 0.19 & 1.36 & 0.49 & 0.42 & 0.24 & 1.08 & 0.39 & 0.34 & 0.19\\
0.39 & 1.08 & 0.22 & 0.26 & 0.49 & 1.36 & 0.28 & 0.33 & 0.39 & 1.08 & 0.22 & 0.26\\
0.34 & 0.22 & 1.08 & 0.14 & 0.42 & 0.28 & 1.36 & 0.18 & 0.34 & 0.22 & 1.08 & 0.14\\
0.19 & 0.26 & 0.14 & 1.08 & 0.24 & 0.33 & 0.18 & 1.36 & 0.19 & 0.26 & 0.14 & 1.08\\
\hline
1.07 & 0.38 & 0.33 & 0.19 & 1.08 & 0.39 & 0.34 & 0.19 & 1.36 & 0.49 & 0.42 & 0.24\\
0.38 & 1.07 & 0.22 & 0.26 & 0.39 & 1.08 & 0.22 & 0.26 & 0.49 & 1.36 & 0.28 & 0.33\\
0.33 & 0.22 & 1.07 & 0.14 & 0.34 & 0.22 & 1.08 & 0.14 & 0.42 & 0.28 & 1.36 & 0.18\\
0.19 & 0.26 & 0.14 & 1.07 & 0.19 & 0.26 & 0.14 & 1.08 & 0.24 & 0.33 & 0.18 & 1.36
\end{array}\right)$$

$$\widehat{\mathbf{\Sigma}}_4=\left(\begin{array}{cccc|cccc|cccc}
1.40 & 0.54 & 0.52 & 0.31 & 1.09 & 0.44 & 0.55 & 0.19 & 1.05 & 0.35 & 0.65 & 0.23\\
0.54 & 1.37 & 0.61 & 0.70 & 0.27 & 0.98 & 0.56 & 0.48 & 0.37 & 1.11 & 0.60 & 0.47\\
0.52 & 0.61 & 2.00 & 0.34 & 0.65 & 0.55 & 1.80 & 0.21 & 0.77 & 0.61 & 1.88 & 0.27\\
0.31 & 0.70 & 0.34 & 1.04 & 0.24 & 0.63 & 0.23 & 0.63 & 0.21 & 0.64 & 0.30 & 0.70\\
\hline
1.09 & 0.27 & 0.65 & 0.24 & 1.63 & 0.52 & 0.75 & 0.21 & 1.22 & 0.28 & 0.81 & 0.19\\
0.44 & 0.98 & 0.55 & 0.63 & 0.52 & 1.19 & 0.56 & 0.56 & 0.42 & 1.01 & 0.60 & 0.52\\
0.55 & 0.56 & 1.80 & 0.23 & 0.75 & 0.56 & 2.14 & 0.16 & 0.82 & 0.61 & 1.93 & 0.20\\
0.19 & 0.48 & 0.21 & 0.63 & 0.21 & 0.56 & 0.16 & 0.75 & 0.16 & 0.50 & 0.18 & 0.61\\
\hline
1.05 & 0.37 & 0.77 & 0.21 & 1.22 & 0.42 & 0.82 & 0.16 & 1.48 & 0.47 & 0.91 & 0.22\\
0.35 & 1.11 & 0.61 & 0.64 & 0.28 & 1.01 & 0.61 & 0.50 & 0.47 & 1.30 & 0.61 & 0.53\\
0.65 & 0.60 & 1.88 & 0.30 & 0.81 & 0.60 & 1.93 & 0.18 & 0.91 & 0.61 & 2.19 & 0.24\\
0.23 & 0.47 & 0.27 & 0.70 & 0.19 & 0.52 & 0.20 & 0.61 & 0.22 & 0.53 & 0.24 & 0.90
\end{array}\right).$$

The estimates show that a space $\times$ time interaction exists as the spatial covariance pattern (among the four caudate radii) changes across the three time points (the 3 $4 \times 4$ blocks along the diagonal of $\widehat{\mathbf{\Sigma}}_4$). However, the separable LEAR model does provide a reasonable approximation to the completely unstructured model with $78 - 5 = 73$ fewer parameters. With $\max_i(t_i s_i) = 147$, the separable LEAR model has $10878 - 5 = 10873$ fewer parameters than a completely unstructured model (for which convergence is currently impossible). Thus, the use of the Kronecker product LEAR model seems acceptable given the statistical and computational benefits it confers.

An alternative approach would be to develop a nonseparable doubly LEAR covariance structure or attempt to embed LEAR structures within an already developed nonseparable framework like that of Fonseca and Steel (2011). Nonseparable techniques have their own unique limitations as discussed in Fonseca and Steel (2011). Examination of these approaches for this context will be the focus of future work.

## 7. Discussion

The Kronecker product LEAR correlation model allows modeling and understanding two factor specific correlation patterns. Excellent analytic and numerical properties make the structure especially attractive for High Dimension, Low Sample Size settings that are common in longitudinal medical imaging and various kinds of longitudinal "-omics" data. The structure is able to model a wide variety of correlation patterns with just four parameters. Analysis of the caudate data illustrates the interpretability of the model in a complex context.

The subset aLRT approach provides a statistically reasonable approach to test the validity of the separability assumption when the standard tests do not apply. More specifically, the diagonal approach we take parsimoniously covers a large fraction of the covariance space and is especially useful in situations like ours where most of the information is contained along diagonal blocks (i.e., correlation dies out along off-diagonals). As evidenced by the simulation results, the subset aLRT is a powerful test that also controls test size with careful selection of the number of significant subtests needed for overall significance.

Future work examining other covariance subspace sampling techniques will prove useful given the contextual nature of the problem. For example, conducting subtests based on random (as opposed to diagonal) subsets of the covariance matrix affords a method amenable to all covariance patterns regardless of whether most of the information is contained along the diagonal blocks. However, the approach may be more computationally intensive and less efficient when most of the information is in the middle of the matrix.

An assessment of model fit and inference accuracy in higher dimensional settings is a priority for future research on Kronecker product LEAR correlation models. Also, introducing a nonstationary Kronecker product LEAR correlation or variance structure may prove extremely useful in neuroimaging since the variability of brain characteristics tends to change over time. Comparing the implementation of the Kronecker product LEAR structure with various modeling and estimation methods will prove valuable. For data that have within-subject correlations induced by three or more factors, as in longitudinal imaging data represented via the m-rep method (Pizer et al. (2002) has details), the generalization of the Kronecker product LEAR correlation model to $F$ repeated factors would be beneficial.

**Acknowledgements**

An earlier version of this manuscript can be found at arxiv.org (Simpson et al., arXiv:1010.4471v1 [stat.AP]). Keith Muller's support includes NIDCR R01 DE020832, R01 DE020832-01A1S1, U54 DE019261, NICHD P01 HD065647, NCRR 3UL1RR029890-03S1, NHLBI R01 HL091005, NIAAA R01 AA016549, NIDA R01 DA031017, and NIDDK R01 DK072398. Sean Simpson's support includes the Translational Scholar Award from the Translational Science Institute of Wake Forest School of Medicine and NIBIB K25 EB012236-01A1.

Table 1. AIC values for all combinations of factor specific correlation models

| | **Initial Caudate Data Model** | | | **Final Caudate Data Model** | | |
|---|---|---|---|---|---|---|
| | Spatial Model | | | Spatial Model | | |
| Temporal Model | LEAR | DE | AR(1) | LEAR | DE | AR(1) |
| LEAR | – 14,298.1 | – 13,903.3 | – 12,716.7 | – 14,306.8 | – 13,911.7 | – 12,722.2 |
| DE | – 14,295.0 | – 13,900.3 | – 12,713.9 | – 14,303.8 | – 13,908.7 | – 12,719.4 |
| AR(1) | – 10,377.3 | – 9,983.2 | – 8,768.1 | – 10,386.1 | – 9,991.8 | – 8,774.0 |

Table 2. Initial full mean model estimates, standard errors, and p-values

| | LEAR $\otimes$ LEAR | | | AR(1) $\otimes$ AR(1) | | | DE $\otimes$ DE | | |
|---|---|---|---|---|---|---|---|---|---|
| Parameter | Estimate | SE | P-value | Estimate | SE | P-value | Estimate | SE | P-value |
| $\beta_0$ | $-5.0172$ | 0.0461 | $< 0.0001$ | $-4.9220$ | 0.0300 | $< 0.0001$ | $-4.9846$ | 0.0439 | $< 0.0001$ |
| $\beta_1$ | 0.0035 | 0.0408 | 0.9309 | 0.0010 | 0.0265 | 0.9689 | 0.0026 | 0.0389 | 0.9463 |
| $\beta_2$ | $-0.0021$ | 0.0034 | 0.5494 | $-0.0026$ | 0.0022 | 0.2398 | $-0.0022$ | 0.0032 | 0.4934 |
| $\beta_3$ | $-0.0362$ | 0.0384 | 0.3458 | $-0.0397$ | 0.0249 | 0.1108 | $-0.0380$ | 0.0366 | 0.2990 |
| $\beta_4$ | $-0.0094$ | 0.0341 | 0.7825 | $-0.0129$ | 0.0220 | 0.5603 | $-0.0104$ | 0.0325 | 0.7484 |
| $\beta_5$ | 0.0162 | 0.0538 | 0.7631 | 0.0144 | 0.0347 | 0.6779 | 0.0160 | 0.0513 | 0.7555 |

Table 3. Final Kronecker product LEAR structure correlation model estimates for caudate data

| Factor | Parameter | Estimate | SE |
|---|---|---|---|
| – | $\sigma^2$ | 0.4047 | 0.0045 |
| Time | $\rho_\gamma$ | 0.9915 | 0.0002 |
| | $\delta_\gamma/(D_\gamma - 1)$ | 0.0026 | 0.0012 |
| Space | $\rho_\omega$ | 0.3806 | 0.0108 |
| | $\delta_\omega/(D_\omega - 1)$ | 0.0402 | 0.0039 |

Table 4. Simulated test size and power, $\alpha = 0.05$, 5,000 realizations.

| | | Structured Test | | | | General Test | | | |
|---|---|---|---|---|---|---|---|---|---|
| | | Test size × 100 | | Power × 100 | | Test size × 100 | | Power × 100 | |
| $N$ | # sig[1] | subset | full | subset | full | subset | full | subset | full |
| 40 | 1 | 8.20 | 0.00 | > 99.99 | > 99.99 | 2.12 | 0.02 | 96.68 | > 99.99 |
| | 2 | 1.10 | – | > 99.99 | – | 0.12 | – | 77.46 | – |
| | 3 | 0.16 | – | 99.96 | – | 0.00 | – | 38.88 | – |
| | 4 | 0.02 | – | 87.56 | – | 0.00 | – | 9.04 | – |
| 80 | 1 | 7.50 | 0.14 | > 99.99 | > 99.99 | 2.30 | 0.24 | > 99.99 | > 99.99 |
| | 2 | 1.04 | – | > 99.99 | – | 0.08 | – | 99.82 | – |
| | 3 | 0.12 | – | > 99.99 | – | 0.00 | – | 90.96 | – |
| | 4 | 0.02 | – | 99.6 | – | 0.00 | – | 54.98 | – |
| 120 | 1 | 7.38 | 0.68 | > 99.99 | > 99.99 | 2.10 | 0.96 | > 99.99 | > 99.99 |
| | 2 | 0.90 | – | > 99.99 | – | 0.12 | – | > 99.99 | – |
| | 3 | 0.12 | – | > 99.99 | – | 0.02 | – | 99.44 | – |
| | 4 | 0.00 | – | > 99.99 | – | 0.02 | – | 84.60 | – |
| 160 | 1 | 8.12 | 1.26 | > 99.99 | > 99.99 | 2.04 | 1.68 | > 99.99 | > 99.99 |
| | 2 | 1.06 | – | > 99.99 | – | 0.06 | – | > 99.99 | – |
| | 3 | 0.06 | – | > 99.99 | – | 0.00 | – | 99.98 | – |
| | 4 | 0.00 | – | > 99.99 | – | 0.00 | – | 96.60 | – |
| 200 | 1 | 7.46 | 1.40 | > 99.99 | > 99.99 | 2.34 | 1.90 | > 99.99 | > 99.99 |
| | 2 | 0.98 | – | > 99.99 | – | 0.16 | – | > 99.99 | – |
| | 3 | 0.16 | – | > 99.99 | – | 0.00 | – | > 99.99 | – |
| | 4 | 0.02 | – | > 99.99 | – | 0.00 | – | 99.38 | – |

[1]# (out of $c = 4$) of significant p-values required for overall significance for the subset aLRT.

Table 5. Simulated test size and power by # (out of $c = 10$) of significant p-values required for overall significance, $\alpha = 0.05$, 5,000 realizations, $\max_i(s_i) = s = 21$, $\max_i(t_i) = 7$, $\text{med}_i(t_i) = 3$.

| | | **Structured Test**[1] | | **General Test**[2] | |
|---|---|---|---|---|---|
| $N$ | # sig | Test size × 100 | Power × 100 | Test size × 100 | Power × 100 |
| 240 | 1 | 35.60 | > 99.99 | 84.18 | > 99.99 |
| | 2 | 13.80 | > 99.99 | 65.92 | > 99.99 |
| | 3 | 4.92 | > 99.99 | 45.52 | > 99.99 |
| | 4 | 1.40 | > 99.99 | 28.24 | > 99.99 |
| | 5 | 0.26 | > 99.99 | 14.24 | > 99.99 |
| | 6 | 0.02 | > 99.99 | 5.92 | > 99.99 |
| | 7 | 0.00 | > 99.99 | 1.94 | > 99.99 |
| 280 | 1 | 21.30 | > 99.99 | 76.04 | > 99.99 |
| | 2 | 5.52 | > 99.99 | 54.80 | > 99.99 |
| | 3 | 1.54 | > 99.99 | 33.66 | > 99.99 |
| | 4 | 0.36 | > 99.99 | 17.50 | > 99.99 |
| | 5 | 0.04 | > 99.99 | 7.62 | > 99.99 |
| | 6 | 0.02 | > 99.99 | 3.02 | > 99.99 |
| | 7 | 0.00 | > 99.99 | 0.74 | > 99.99 |
| 320 | 1 | 14.46 | > 99.99 | 73.64 | > 99.99 |
| | 2 | 3.00 | > 99.99 | 49.60 | > 99.99 |
| | 3 | 0.56 | > 99.99 | 29.22 | > 99.99 |
| | 4 | 0.10 | > 99.99 | 14.42 | > 99.99 |
| | 5 | 0.02 | > 99.99 | 6.04 | > 99.99 |
| | 6 | 0.00 | > 99.99 | 1.76 | > 99.99 |
| | 7 | 0.00 | > 99.99 | 0.38 | > 99.99 |

[1] $\nu = 100$ (degrees of freedom) for each subset test (10 subset tests)
[2] $\nu = 75$ (degrees of freedom) for each subset test (10 subset tests)

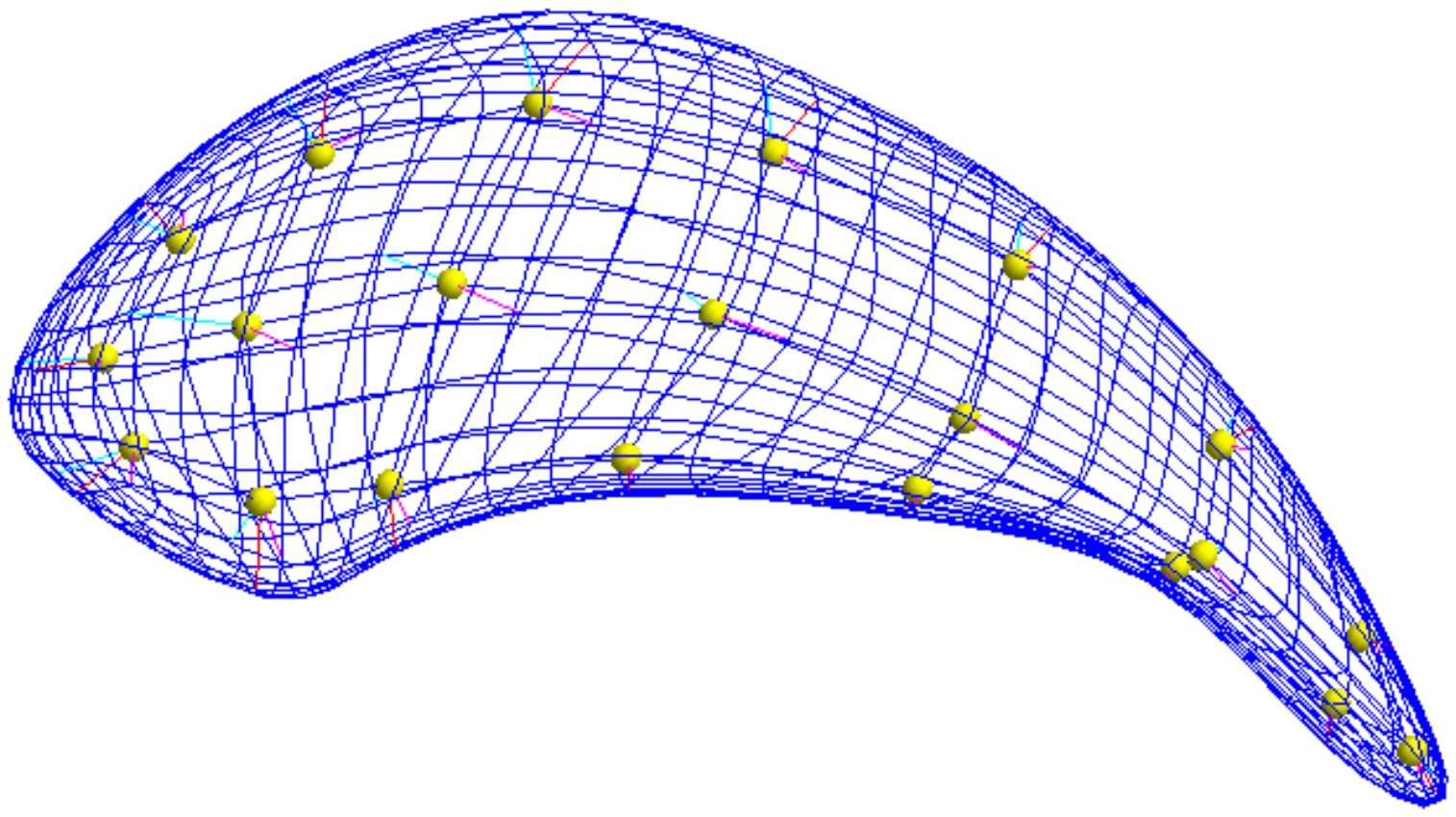

Figure 1. M-rep shape representation model of the caudate.

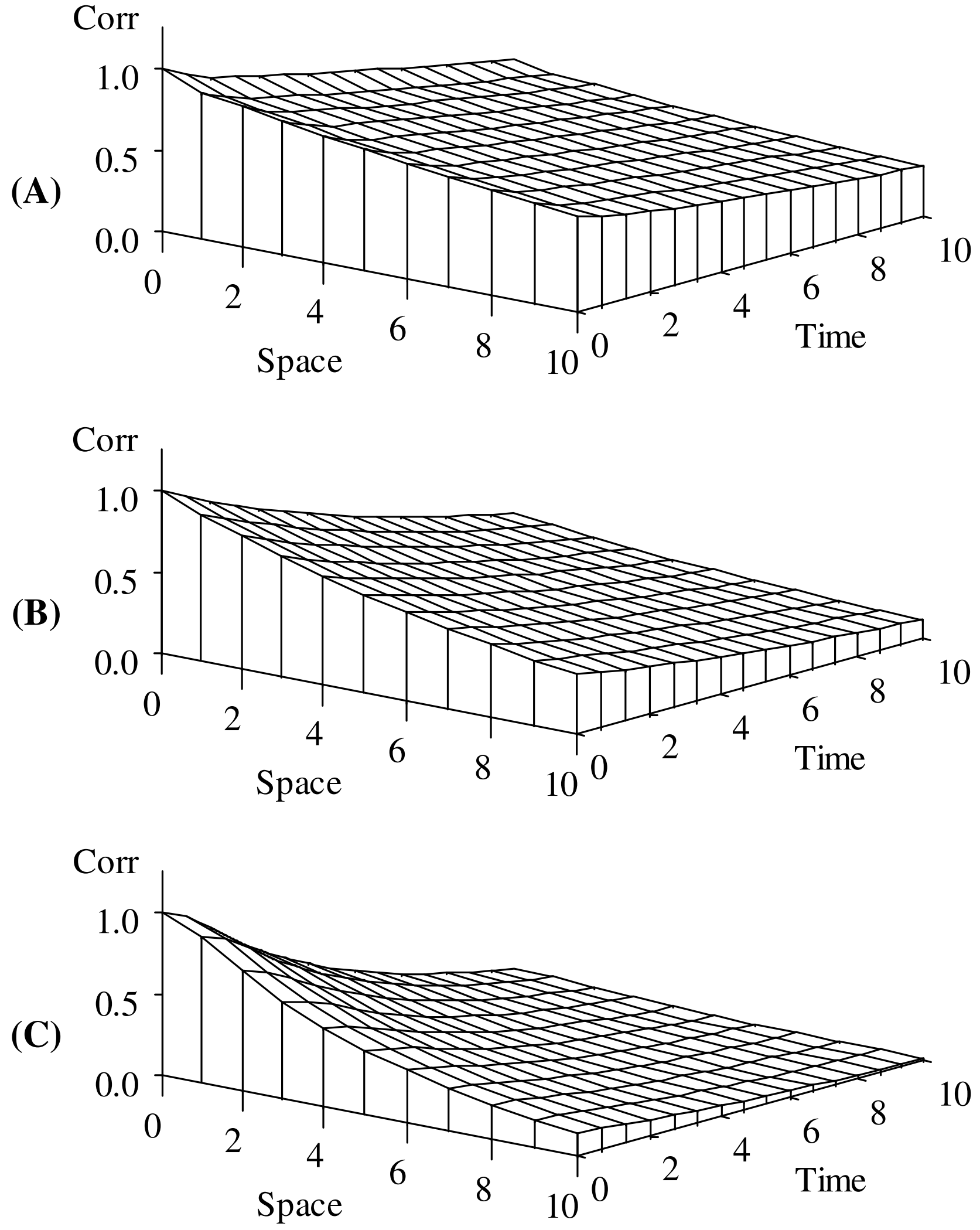

Figure 2. Plot of correlation as a function of spatial and temporal distance.

**(A)** Both factors have a decay slower than AR(1) with $\{\rho_\gamma = 0.9, \rho_\omega = 0.9\}$ and $\{\delta_\gamma/(d_{t;\max} - d_{t;\min}) = 0.5, \delta_\omega/(d_{s;\max} - d_{s;\min}) = 0.5\}$.

**(B)** Both factors have an AR(1) decay with $\{\rho_\gamma = 0.9, \rho_\omega = 0.9\}$ and $\{\delta_\gamma/(d_{t;\max} - d_{t;\min}) = 1, \delta_\omega/(d_{s;\max} - d_{s;\min}) = 1\}$.

**(C)** Both factors have a decay faster than AR(1) with parameters $\{\rho_\gamma = 0.9, \rho_\omega = 0.9\}$ and $\{\delta_\gamma/(d_{t;\max} - d_{t;\min}) = 2, \delta_\omega/(d_{s;\max} - d_{s;\min}) = 2\}$.

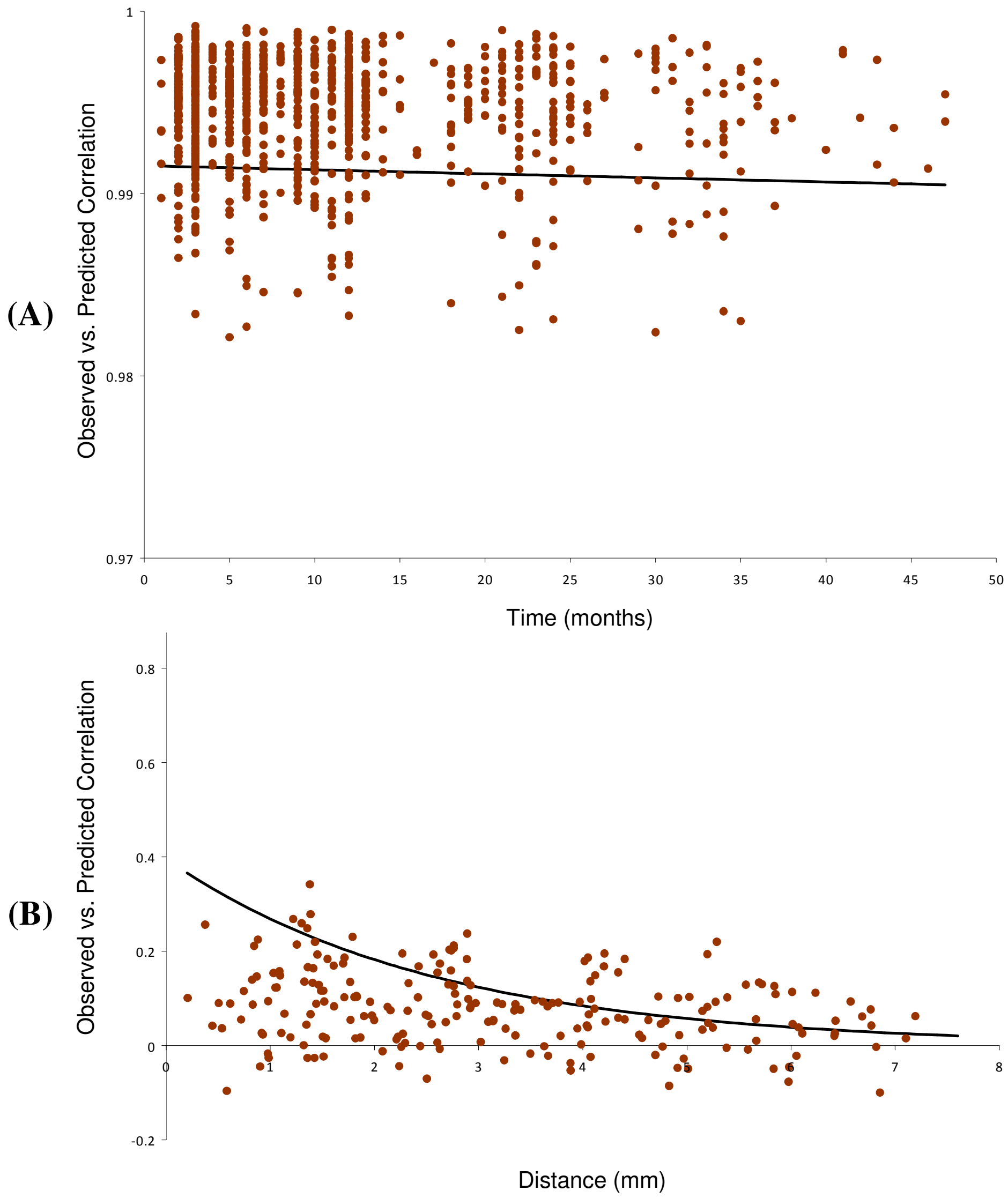

Figure 3. Observed (dots) vs. predicted (curve) correlation: **(A)** as a function of the time between images; **(B)** as a function of the distance between radius locations.

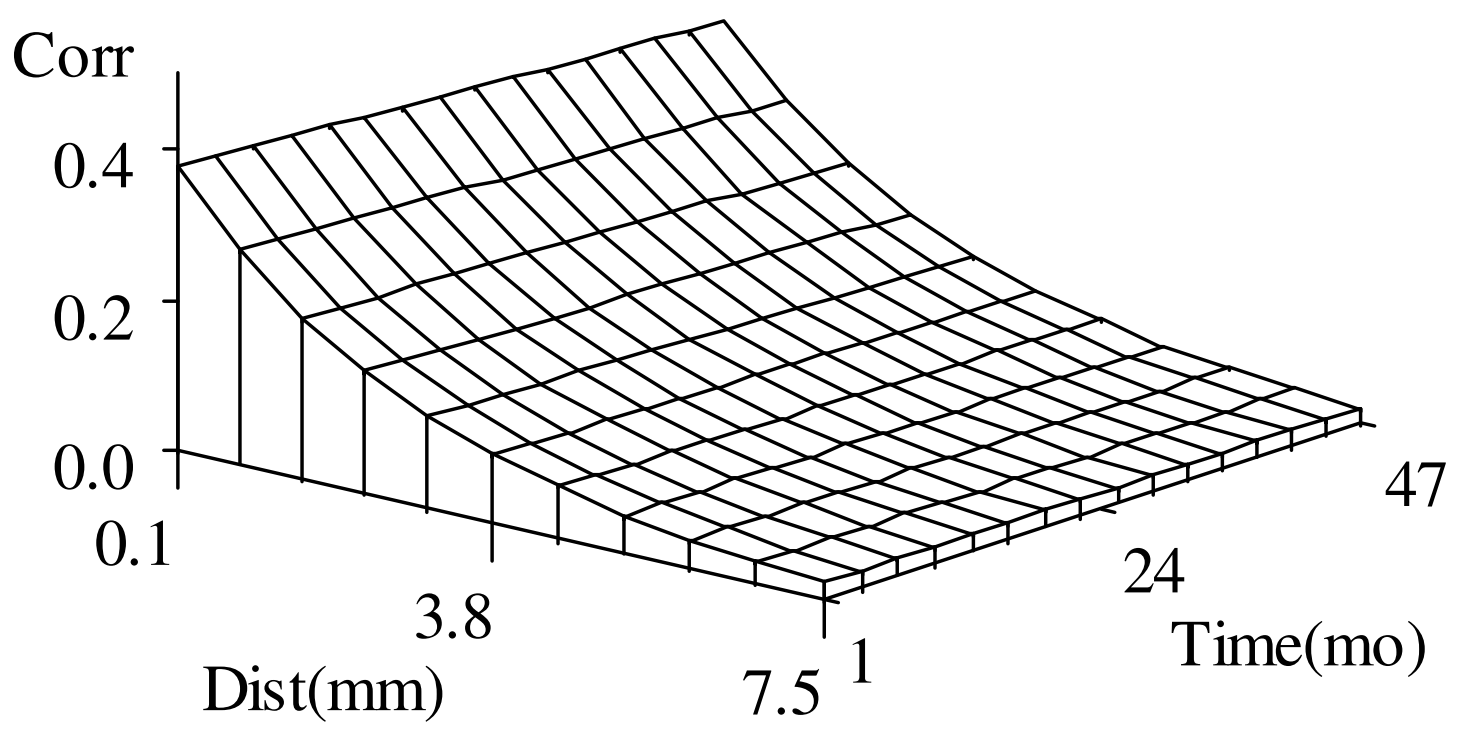

Figure 4. Predicted correlation as a function of the distance between radius locations and time between images.